# Commuter networks and community detection: a method for planning sub regional areas

Andrea De Montis[(1,4)], Simone Caschili[(2)], Alessandro Chessa[(3,4,5)]

[(1)]*Dipartimento di Ingegneria del Territorio, Sezione Costruzioni e Infrastrutture*
*Università degli Studi di Sassari Via De Nicola – 07100 Sassari, Italy,*
*e-mail: andreadm@uniss.it*

[(2)]*Dipartimento di Ingegneria del Territorio, Sezione di Urbanistica,*
*via Marengo 3 - 09127 Cagliari, Italy*
*e-mail: scaschili@unica.it*

[(3)]*Dipartimento di Fisica, Università degli Studi di Cagliari*
*Cittadella Universitaria di Monserrato, 09042 Monserrato, Italy*
*e-mail: alessandro.chessa@dsf.unica.it*

[(4)] *Linkalab, Complex Systems Computational Laboratory, 09129 Cagliari, Italy*

[(5)]*Istituto Sistemi Complessi - CNR, UOS "Sapienza", Piazzale A. Moro 2, 00185-Rome, Italy*

Abstract

A major issue for policy makers and planners is the definition of the "ideal" regional partition, i.e. the delimitation of sub-regional domains showing a sufficient level of homogeneity with respect to some specific territorial features. In Sardinia, the second major island in the Mediterranean sea, politicians and analysts have been involved in a 50 year process of identification of the correct pattern for the province, an intermediate administrative body in between the Regional and the municipal administration. In this paper, we compare some intermediate body partitions of Sardinia with the patterns of the communities of workers and students, by applying grouping methodologies based on the characterization of Sardinian commuters' system as a complex weighted network. We adopt an algorithm based on the maximization of the weighted modularity of this network to detect productive basins composed by municipalities showing a certain degree of cohesiveness in terms of commuter flows. The results obtained lead to conclude that new provinces in Sardinia seem to have been designed -even unconsciously- as labour basins of municipalities with similar commuting behaviour.

# 1 Introduction and motivation

A major issue for analysts and planners has often been and still remains the assessment of the "ideal" spatial unit, i.e. a suitable geographical domain to support master planning on areas so complex that deserve a division in sub areas constituted by similar elements. As a consequence, many scholars have reflected on the nature and scope of the appropriate territorial unit and many others have constructed methods able to support the research of homogeneous sub-areas, under a variety of perspectives. Recently, complex network analysis (CNA) has provided a powerful methodology to study large systems –i.e. systems constituted by a large number of elements interlaced by numerous connections. Under this general framework, many scholars have investigated on methods able to detect communities made up by cliques of nodes –i.e. subsets of nodes and their mutual connections- that display a certain degree of similarity.

Starting from this background remarks, in this paper the authors refer to the assessment of inter-municipal commuting basins in the island of Sardinia, Italy, by adopting community detection tools belonging to the family of network analysis based algorithms. They aim at comparing the community structure obtained to relevant administrative and cultural subdivisions (i.e. the province an intermediate body in the Italian administrative system ). We test the hypothesis that the new distribution of provincial bodies, introduced by the State in 2003, correspond -to a large extent- to the pattern of commuting basins.

The argument is developed as follows. In the next section, a brief state of the art is reported on the concepts and methods useful for assessing appropriate spatial units. In section 3, a synthetic state of the art is also described on CNA and the advances in the field of community detection on networks. In section 4, the community detection method selected in this case is described, while in section 5 it is applied to the evaluation of productive commuter basins of Sardinia through time. Section 6 presents the conclusions of this paper and proposes some outlook remarks.

# 2 Regionalization in urban, land, and regional planning: some theoretical developments

Regionalization -i.e. the assessment of appropriate territorial units- has been one of the focal point of the researches of many scholars that have elaborated methodologies and applications under a variety of perspectives, ranging from productive and administrative basins to service catchments population areas.

In a pure planning and decision making perspective, Archibugi (1994) in general reflects on the proper scale and functions to be attributed to the appropriate spatial units. In particular, starting from the Italian case, he proposes a methodology to define urban mobility integrated basins (Archibugi, 1993). In these investigations, Archibugi studied the impact of the urban agglomerates over the surrounding ecosystems by introducing the concept of basin, whose footprint is larger than the sole city.

In a perspective closer to geography and regional analysis, Berry has set the ground for general concepts, such as regionalization and regional character (Berry, 1964), instrumental tools, such as the functional economic areas (Berry, 1966 and 1968), and operational principles, such as cluster and multivariate analysis (Berry, 1967). Fox's contribution to this field of analysis concentrates both on functional economic areas (Fox, 1967) and on methods suitable for locating and characterizing these areas (Fox, 1974). Geographers have devoted a great interest in defining features that distinguish a particular region and allowing its delimitation over the Earth. This can be referred to the attention they have paid for the broader scientific tradition of studies that provide "a description and interpretation of a region or place by considering a series of attributes of that place" (Wheeler, 1986, p. 54). This attention is documented during the first decades of the last century: according to Davis (1924), geographers are first of all regionalists.

As geographer interested in a more statistical perspective, Fischer has contributed to the early reception and elaboration of taxonomic studies in the domain of geography. He focused on methods able to group and classify objects according to their measured social, geographic, and economic properties (Fischer, 1979, 1982, 1984). Under this approach, many scholars have studied the possibility to apply multivariate and, in particular, factor analysis, as algorithms able to delimitate regions and interpret patterns (Garrison and Marble, 1963; Davies, 1980; Spence and Taylor, 1970). The advances in statistics demonstrate a vivid interest for assessing regionalization algorithms able to take into account uncertainty and risk. Under this point of view, fuzzy sets theory provides us with many significant contributions (Iacovacci, 1995; Leung, 1988; Rolland-May, 1985; Zadeh, 1977).

Noronha and Goodchild (1992) have put the emphasis on spatial analysis and its possible role in functional regionalization. Starting from a review of studies on pioneering research about computer-based methods for delimitating regions, they propose a distribution of US districts of commuting students taking into account selective spatial interaction among them.

In the perspective of geographic information scientists, many authors stress that regionalization algorithms implying information intensive procedures should be conceived and tailored on the data set available and, vice versa, spatial data architecture may affect the pattern of functional regions obtained (Batty and Sammons, 1978; Masser and Brown, 1975; Semple et al, 1972). A number of authors discussed and proposed methods –i.e. exploratory spatial data analysis (ESDA)- that may be conceived as modules of GIS useful in many fields ranging from urban and regional model to public health (Anselin and Getis, 1992; Anselin, 1998; Batty and Xie, 1994; Goodchild et al., 1992; Rushton, 2003; Symanzik et al., 1996).

In the next section, a brief state of the art is proposed on the principles of community detection over systems conceived as networks.

## 3    Complex networks and community detection

Network theory begins with the elaboration of graph theory by Euler[1]. This was developed and expanded in the 1960's by Erdős and Rényi (1959, 1960). In a few seminal works they studied the structure of random graphs, where each pair of nodes in the graph is linked with a certain probability p. Afterwards, the availability of ever larger data sets and the parallel explosion of computer processing power allowed Complex Network Analysis (CNA) to be systematically and intensively applied to the study of very large networks (Pastor Satorras and Vespignani, 2004; Albert and Barabàsi, 2002). These networks have a very large number of nodes, N, when compared to the average degree value (N >> <k>), where k represents the number of first neighbour nodes connected to a given node *i*.

At the end of the last century an interest in Network Theory revamped with the work of Watts and Strogatz (1998). They first proposed a simple network model to explain a very well known social spreading phenomenon called small world effect (Milgram, 1967). These networks model social interactions in a regular one-dimensional lattice. A random rewiring process is introduced in order to generate the small-world topology that is characterized by an average shortest path length L that scales very slowly with the number of nodes N (L~logN). Another important property, that has been found in many natural and technological networks, is that the probability distribution of the degree P(k) does not provide any characteristic value in presence of a diverging measure of the statistical fluctuations (variance). Conversely, networks with a power law trend for the probability distribution of the degree k are said to be scale-free because they have invariant statistical properties over the entire range of degree values. The presence of heavy tails in the probability distribution curve is also often a sign of a non-negligible probability to encounter very large degree nodes, the "hubs" of these networks. A popular mechanism for explaining the growth of these special networks is the so-called preferential attachment (Barabàsi and Albert, 1999).

---

[1] One of the greatest mathematicians ever: applying graph theory for the first time in his Solutio problematis ad geometriam situs pertinentis (1736) he demonstrated that it was impossible to complete a leisurely walk of the city of Königsberg by crossing its seven bridges only once (after Caldarelli, 2007).

Recent developments in CNA have regarded the analysis of weighted networks, where a quantity -the weight- is attributed to each edge (Barrat et al, 2004; Barthélemy et al, 2005). In these approaches, modellers adopt a generalization of the measures developed to study the purely topological networks' characteristics. Other relevant results have arisen from the analysis of the interplay between weighted measures and topology. Super proportional correlations have been identified and showed that the flow of information in a complex network cannot be fully explained solely by its topology.

CNA has been applied to both man-made and natural systems. Apart from computer simulations, CNA provides insights into a wide range of topics such as food webs, human interactions, the Internet, the World Wide Web, the spread of diseases, population genetics, genomics and proteomics. In each of these cases one starts by inspecting recurrent structures embedded in complex systems characterized by non-identical elements (the nodes) connected through different kinds of interactions (the edges). For a review of these applications, see Albert and Barabàsi (2002) and Newman (2003).

Recently, a number of regional scientists have adopted CNA for modelling urban (Batty, 2001, 2008a, 2008b; Jiang and Claramunt, 2004; Barthélemy and, Flammini, 2008 and 2009), regional, social and economic systems (Latora and Marchiori, 2003; Schintler et al, 2005). Many authors have attempted to extend the analysis beyond topology and traffic by inspecting the influence of geographical space on the properties of the network (Gorman and Kulkarni, 2004; Gastner and Newman, 2004; Crucitti et al, 2006; Campagna et al, 2007).

Of very special interest are those works, that have applied CNA to the study of commuting. Patuelli et al (2007) analysed the topology of the German commuting network. De Montis et al (2007) used a weighted network approach to analyse inter-municipal commuting in the Italian region of Sardinia, the second largest island in the Mediterranean.

Very promising research on complex network theory attains the detection of communities (for a review, see Fortunato, 2009). A community in a network may be defined as a set of nodes that present a high number of internal links and few links toward the nodes belonging to other communities. So, communities may correspond to groups of pages of the World Wide Web dealing with related topics (Flake et al., 2002), functional modules such as cycles and pathways in metabolic networks (Guimerà and Amaral, 2005; Palla et al., 2005), to groups of related individuals in social networks (Girvan and Newman, 2002; Lusseau and Newman, 2004), and compartments in food webs (Pimm, 1979; Krause et al., 2003).

In a more general perspective, Fortunato and Castellano (2009) group the definitions of communities in three main categories: local, global and based on vertex similarities. In local definitions, the attention is focused on the vertices of the sub-graph under investigation and on its immediate neighbourhood, disregarding the rest of the graph. Global definitions include communities as structural units of the graph, so it is reasonable to think that their distinctive features can be recognized if one analyses a sub graph with respect to the graph as a whole. Definitions based on vertex similarity select communities as groups of vertices, which are similar to each other. A quantitative criterion is chosen to evaluate the similarity between each pair of vertices, connected or not.

Community detection over a network is affected by the characteristics of the system at hand. While in random networks a heterogeneous pattern overall emerges, in real networks small-world and scale free effects often lead to statistically homogeneous structures.

Furthermore, the distribution of edges is not only globally, but also locally homogeneous, with high concentrations of edges within special groups of nodes, and low concentrations between these groups (Fortunato and Castellano, 2009). Traditionally, community detection in graphs aims at identifying the modules only based on the topology. The problem has a long tradition and it has appeared in various forms in several disciplines. New advances also propose the study of detection of communities in weighted networks where not only the topology influences the shaping of clusters but also the weight of each link (in terms of resilience or strength of interaction between nodes).

In particular, community detection may become a complex activity if we consider systems comprehending several nodes and links. Communities tend to overlap each other showing some nodes in common throughout the network. Another important issue in the case of large networks regards graphs whose vertices have various levels of organization. Communities can display an internal structure, i.e. they can contain smaller communities, which can in turn include other communities. In this case one says that the community structure is hierarchical.

Many authors have proposed methods and algorithms to detect communities in networks. The literature indicates three main methods: divisive algorithms, optimization methods, and spectral methods. Other approaches that do not fit in the clusters above are the following: Q-state, Potts model, clique percolation, random walk, Markov cluster algorithm, maximum likelihood, and L-shell method (Fortunato and Castellano, 2009).

With respect to the literature overview presented so far, it is possible to envisage an interest of analysts and planners for network based community detection methods. These tools are able to depict patterns starting from the analysis of similarities among simple elements (the nodes) intertwined in a known topology. This important goal can be achieved, since those methods convey additional information, with respect to the traditional clustering methods described in section 2.

In the next section, we illustrate the Louvain algorithm as proposed by Blondel et al. (2008). Therefore we apply it to detect territorial clusters shaped by commuting in the Italian region of Sardinia.

## 4   A heuristic method to detect clusters in a regional commuting network

In the panorama of studies developed during the last half-decade in networks' community detection, many scholars focused on the performance of the various methods, while a few authors proposed extensive practical applications.

In this paper, we inspect the community structure provided by the movements of commuters in the region of Sardinia. We propose the application of the Louvain algorithm, proposed by Blondel et al. (2008), an optimisation method based on the maximisation of an objective function called modularity (Newman and Girvan, 2004), defined as follow for the case of weighted networks:

$$Q_w = \frac{1}{2W} \bullet \sum_{ij}\left(w_{ij} - \frac{s_i s_j}{2W}\right) \bullet \delta(c_i, c_j) \qquad (1)$$

where $w_{ij}$ is the weight associated to the edge connecting the node $i$ and the node $j$, $s_i = \sum_j w_{ij}$ (node strength) is the sum of the weights of the edges attached to the node $i$, $W = \frac{1}{2}\sum_{ij} w_{ij}$ is the sum of all the edge weights, and $\delta(c_i, c_j)$ is a function equal to one, when vertices $i$ and $j$ belong to the same community, and to zero otherwise.

The modularity function tries to quantify how good is a community subdivision, among all possible ones, by computing, for a particular subdivision, how many edges there are inside the communities with respect to the number of edges among them. Its values range from -1 to +1. The 0 value occurs when a certain subdivision has no more intra-community edges that one would expect by random chance. A negative value means that there is no advantage in splitting the network in communities and the best solution is one community.

The Louvain algorithm is quite interesting, since it allows one to successfully approach two critical issues of optimisation methods: detecting communities in large networks in a short time and taking into account hierarchical community structure. The algorithm is based on an iterative process whose steps are shown in Table 1.

| Steps | Task |
|---|---|
| 1 | Each node is assigned to a unique single community |
| 2 | Neighbour nodes of each target node are preferentially included in the same community if the variation of the modularity ($\Delta Q_w$) is positive |
| 3 | This aggregation process proceeds until the modularity function $Q_w$ reaches a maximum |
| 4 | A new network is then built whose nodes correspond to the communities obtained in step 3; each link connecting a pair of communities is featured by a weight equal to the sum of the weights of the external links originally between them. The internal links are represented by a self-loop, whose weight is equal to the sum of their internal weights |
| 5 | Step 1 applies to the last network |

Table 1 Overview of the iterative process supporting the algorithm by Blondel et al (2008).

This algorithm may be used for both weighted and un-weighted networks.

The $\Delta Q$ function, introduced in step 2 above, measures the level of performance of the partition associated to the displacement of a node from a community $C$ to another.

The algorithm naturally incorporates a notion of hierarchy, as communities of communities are built during the process (Blondel et al., 2008). This theoretical construction reminds the scale invariance and self-similarity properties of complex systems (Song et al., 2005). Moreover the number of communities at each hierarchical level emerges naturally from the algorithm and it has not to be imposed in the beginning, as in other clustering approaches.

In the next section, an application of this method is reported, as it enables one to investigate communities emerging in the system of inter-municipal commuting of the island of Sardinia.

## 5  Comparing commuter basins and administrative bodies

In this section, we present an application of the community detection methodology described in section 4 to support an ex post discussion about the recently instituted new provinces of the Italian island of Sardinia (RAS, 2001 and 2003). The initial assumption is that an administrative subdivision (i.e. the provincial body) should represent adequately also an homogeneous mobility basin –i.e. a community of municipalities with similar commuters' behaviour. We aim at testing our results by comparing the territorial partitions provided by the Louvain method to some public administration subdivisions that were instituted in the last fifty years in Sardinia.

In this case study, a comparison between different partitions of communities will be developed, as follows. First, we report on the evolution of administrative units in Italy and Sardinia – i.e. the fifty-year evolution of the province, the Italian intermediate administrative body. Second, commuter basins –i.e. homogeneous communities of municipalities with a similar commuter behaviour- will be extracted, via the algorithm set by Blondel et al. (2008). Third, the sets of partitions above will be confronted, by adopting a quantitative method based on the adjusted Rand index (Hubert and Arabie, 1985).

*5.1 The identification of functional sub regional units in Sardinia*

In modern states, the whole national territory is usually subdivided in homogeneous sub units. In the European Union, there are in-between units such as municipalities (the basic administrative gathering communities NUTS 3 or LAU 1 units) and regions (or similar entities at a state level - NUTS 2 units). Historically in Europe those in-between districts have different names and carry out different tasks: in France "Le departement" dates back to the Napoleonic age, counties are an English tradition, "Regierungsbezirk" in Germany, "Provincie" in Italy, and "Disputaciones" in Spain.

That kind of sub divisions can be identified according to both a normative and an analytical regulation. In fact, they are units with similar spatial and demographic features, corresponding to areas that range from mid to large size. Those units are usually inhabited by a population from 100 to 500,000 citizens that live at a like distance around a big town.

In this contest, we focus on the Italian administrative hierarchical organization as we are going to discuss a community detection network method for the recognition of productive, social and administrative territorial units in Sardinia. Actually, Italy is subdivided into 20 regions (Regione in Italian) that are further divided into 110 provinces (Provincia) and 8,100 municipalities (Comune).

Nevertheless while the regions and municipalities kept a strong configuration since they have been founded, Provincia with changing fortunes and cyclical successes assumed different roles and strategies in the Italian territorial organization. During the 1960s, 1970s and 1980s, the institution of new in-between boards similar to provinces but smaller such as comprensori, comunità montane (mountain community), unità sanitarie (health districts), distretti scolastici (school districts), raised the discussion about which body could better represent and meet the demands of the local communities.

Moreover the administrative devolution brought by these new bodies introduced new problems. For the past fifty years, inspired by the socio-political propulsions of the 1960s, administration of communities has devolved from the central government to local bodies. This shift in power gave more administrative independence to local communities, under the principle that living societies have a higher awareness about their own needs than a distant central body.

The region of Sardinia has gone through periods of devolution and reorganization of local power (De Montis, 1997). In table 2 we report on the four phases that have characterized the devolution in Sardinia from the 1950s to present.

| Phases | Territorial sub-divisions | Time period |
|---|---|---|
| 1 | Piano di Rinascita | 1950s-1960s |
| 2 | Comprensori | 1970s |
| 3 | Profili d'Area | 1990s |
| 4 | New Provinces | 2000's |

Table 2 The phases characterizing the evolution of the intermediate body in Sardinia.

Figure 1 displays a geo-referred visualisations of the above cited units. In the remainder of this paragraph, we will introduce the duties of these administrative units and the reasons that brought the regional administration of Sardinia to set and sometimes even institute them.

The idea to design the "Piano di Rinascita" (the Revival Plan) started to emerge at the turn of the 1950s during the first Sardinian Congress. The plan became effective when national Law 158/1962 was issued. Piano di Rinascita was the first regional act that attempted to implement a strategic planning for the whole region. The aim of Law 158/1962 was to promote a twofold general goal: the economic and social development of Sardinia. In the 1950s the economic gap between Sardinia and the other Italian regions was so vast that it was felt that something had to be done to bring its economy into line with the rest of the country. Piano di Rinascita had several goals including: environmental and building renovation; redevelopement of agriculture; industrial renewal; encouraging and regulating the development of fishing, handicraft, business and tourism sectors.

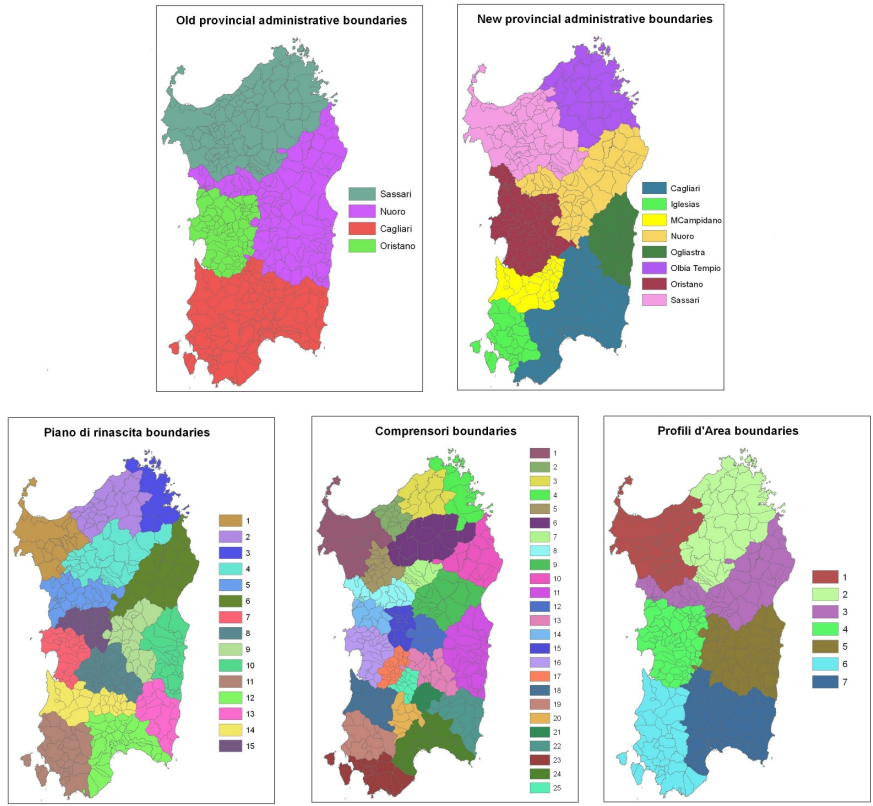

Figure 1: maps of the relevant administrative bodies that we use to test similarities and dissimilarities with cluster network analysis in the regional setting of Sardinia.

The Piano di Rinascita was constructed subdividing the regional area into fifteen homogeneous clusters. These units were selected according to the social and economic framework and the likelihood of improving their conditions. The final goal was to improve the general conditions of the fifteen territorial units by, among other things, encouraging a decrease in unemployment and an increase in the average income.

As a second spatial unit, in Sardinia, Comprensorio units were founded with the regional Law 33/1975 (but never became effective). This law was inspired by principles of devolution and social involvement. Another goal achieved by that law was to apply bottom-up planning with the cooperation of local communities. Law 33/1975 founded twenty-five Comprensorio. Each Comprensorio had the function of drawing up a multi-year plan for: controlling and organizing urban development; organizing and distributing social facilities in the territory; coordinating development in the agricultural sector.

The purpose of the central government was to put into effect the formation of smaller functional sectors more responsive to the local communities and conditions by assigning to Comprensorio functions of land use and urban transformation.

Thirdly, the "Piano di sviluppo" (Regional Development Plan) was approved by the regional planning department in 1987. Its goal was to improve the economic conditions of Sardinia by focusing on the environment. This was a completely new approach. The "Piano di sviluppo" defined seven homogeneous units of action, named "Profili d'Area" and had several aims including: suggesting projects to achieve better economic conditions; providing a democratic instrument to judge utility and coherence of projects with the established territorial goals; judging the agreement of projects with the goals of environmental sustainability.

Profili d'Area were territorial units with set boundaries. Units were determined by economic, social, and environmental factors. The plan's goal was to facilitate projects in areas that would have a shared interest in the project but are not necessarily within the same units.

Turning to the fourth spatial unit introduced above, in the Italian public administration structure, the province represents an intermediate body in between the regions and the municipalities, hence able to transfer with high efficiency the acts and regulations of a high-profile region to the minute world of the municipalities. In the last decades, the number of the provinces has increased at the national level, while in Sardinia it has remained stable for many years. Up until 2003, there were four provinces while Sardinia is currently partitioned in eight provinces, after a fifty-year process where the most appropriate sub regional division has been sought. The institution of the four new provinces has attempted to guarantee greater economic development and apply the principle of subsidiary action, with respect to the central regional administration.

*5.2    Application to the regional commuting network of Sardinia*

According to recent studies (De Montis et al. 2007, 2010, 2011), the inter-municipal commuting system of Sardinia, corresponding to the daily movements of workers and students, can be conceived as a network. For the whole set of municipalities the Italian National Institute of Statistics has issued the origin-destination table (ODT) corresponding to the commuting traffic at the inter-city level. The ODT is constructed on the output of a survey about commuting behaviors of Sardinian citizens. This survey refers to the daily movement from the habitual residence (the origin) to the most frequent place for work or study (the destination). The dataset comprises information both on the transportation means used and the time usually spent for displacement. Hence, ODT data give access to the flows of people regularly commuting among the Sardinian municipalities. In particular we have considered the *external flows i→j* that measure the movements from any municipality $i$ to the municipality $j$. We will take into account the complete flows of individuals (workers and students) that commute throughout the set of Sardinian municipalities by all means of transportation. This data source allows the construction of the Sardinian inter-municipal commuting network (SMCN) in which each node corresponds to a given municipality and the links represent the presence of a non-zero flow of commuters among the corresponding municipalities. We can thus construct the symmetric weighted adjacency matrix **W** in which the elements $w_{ij}$ are computed as the sum of the *i→j* and *j→i* flows between the corresponding municipalities (per day). The elements $w_{ij}$ are null in the case of municipalities *i* and *j* which do not exchange commuting traffic and by definition the diagonal elements are set to zero $w_{ii}=0$. According to the assumption of regular bi-directional movements along the links, the weight matrix is symmetric and the network is described as an undirected weighted graph.

In this way, the Sardinian inter-municipal commuting network (SMCN) is composed by a set of vertices corresponding to the towns, and a set of edges representing the exchange of commuters between two towns.

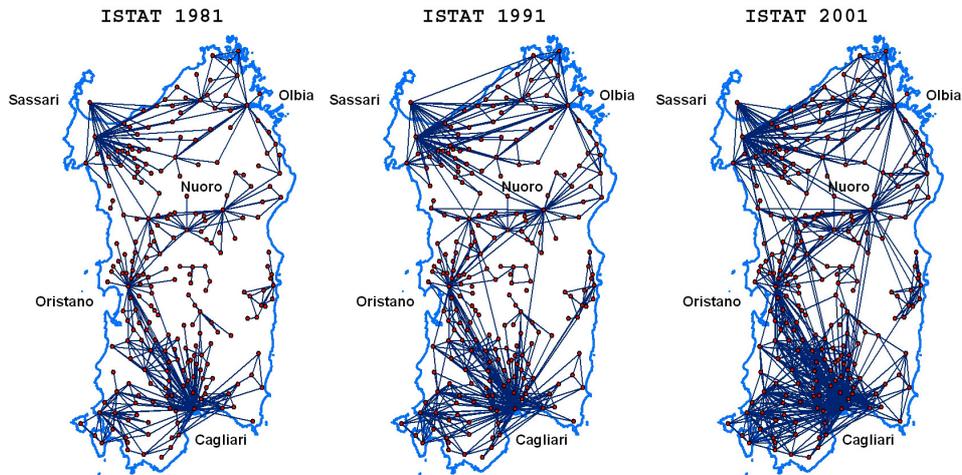

Figure 2 Geo-referred representation of the SMCN in the years 1981, 1991, and 2001.

In the remaining of this section, the application of community detection on the SMCN, represented in Figure 2, is reported for the years 1981, 1991 and 2001. The structure of communities obtained is interpreted and compared with the pattern of relevant administrative bodies, such as the provinces of Sardinia.

The hierarchy of the communities, i.e. an important indication of the granularity of the network structure, may be pictured by a dendrogram that displays a number of layers corresponding to different orders of communities. The dendrogram has three main layers: in layer 1, each node constitutes a community, in layer 2 around 20 communities are detected, and in layer 3 around 8. In Figure 3, we use a dendrogram in order to visualize how the Louvain method clustered the municipalities of the SMCN in 1981. For ease of reading, we report a partial representation of the dendrogram. The diagram clearly shows the hierarchical organization of the network and the number of layers corresponding to different orders of communities.

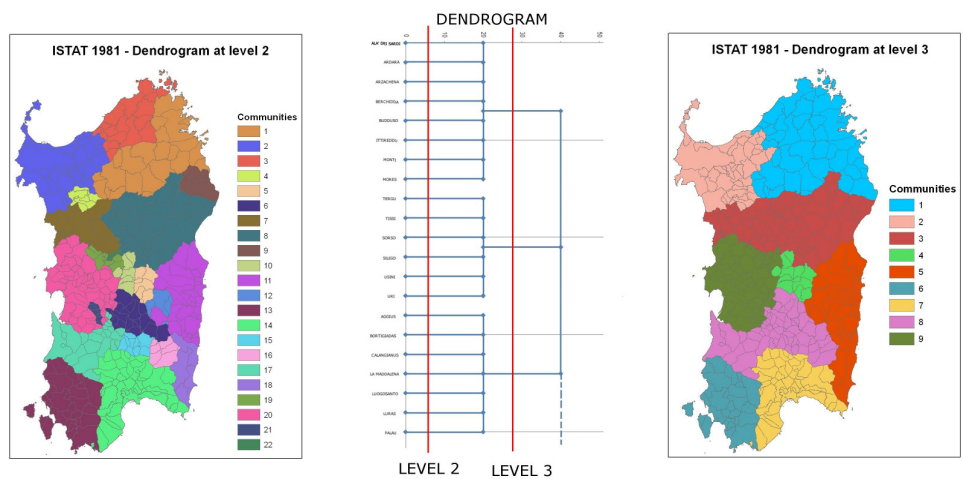

Figure 3: A partial representation of the dendrogram for the ISTAT commuter movements in 1981. First community level belongs to single municipality communities. On the left, the geographical representation of communities at level 2, on the right at level 3.

In Table 3, we summarize and compare the number of communities for the three years (1981, 1991 and 2001) with the administrative units selected (in Section 5.1).

Table 3: number of communities for each dataset in the 20 years period

|  | 1981 | 1991 | 2001 |
|---|---|---|---|
| SMCN layer 2 communities | 22 | 19 | 19 |
| Comprensori |  | 25 |  |
| SMCN layer 3 communities | 9 | 7 | 8 |
| Old provinces |  | 4 |  |
| New provinces |  | 8 |  |
| Piano Rinascita |  | 15 |  |
| Profili d'Area |  | 7 |  |

In the comparison of the basins detected with the relevant divisions of Sardinia, we have selected and confronted layer 3 communities with the provinces, Piano di Rinascita and Profili d'Area, while layer 2 communities with the Comprensori.

In Figure 4, a synthetic map of the overlay between layer 3 communities and old provinces in the years 1981, 1991, and 2001 is reported. It is immediately worth noting that: in 2001 the number of communities mirrors the number of new provinces (eight); in each year the communities are geographically connected and there are no disconnected "islands"; in each year in many cases the boundaries coincide. It is to be stressed that the network we used has no notion of space, since it topologically maps the exchange of commuters regardless of the vicinity of the municipalities. So it is not trivial that the algorithms a posteriori detect sensible and connected geographic regions.

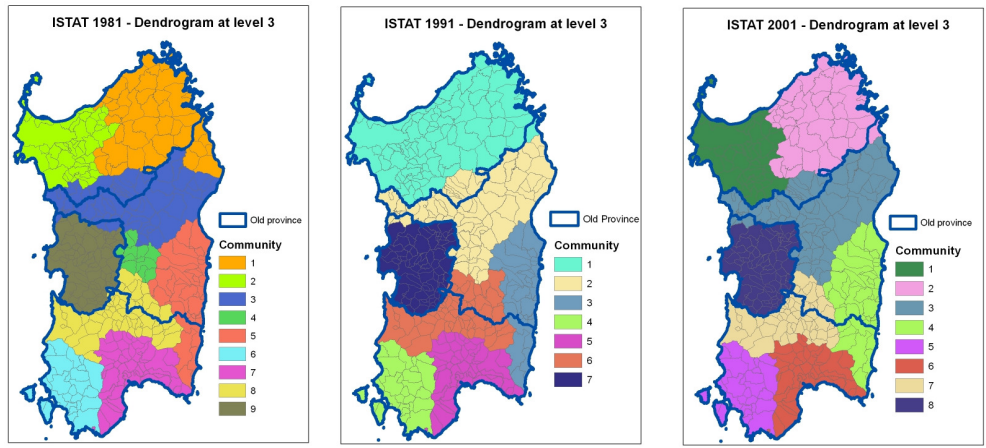

Figure 4 Overlay of layer 3 communities with old provinces in the years 1981, 1991 and 2001.

When we look at the comparison between the boundaries of the new and old provinces and our partitions, a high degree of similarity was detected among the province of Oristano, community 9 (Figure 4 on the left), community 7 (Figure 4 in the middle) and community 8 (Figure 4 on the right). A greater degree of similarity can be detected from the comparison of newer provinces than older ones. Even when reviewing new provinces we can visually detect a high spatial coincidence: the partitions at level 3 communities seem to better reflect the boundaries of the new provinces. In the northern part of Sardinia (local zones called Sassarese and Gallura) provinces are well reproduced by the communities 1 and 2 in Figure 5 for the years 1981 and 2001. A high degree of similarity can be detected for the area of Nuorese (the central zone of Sardinia) with level 3 communities.

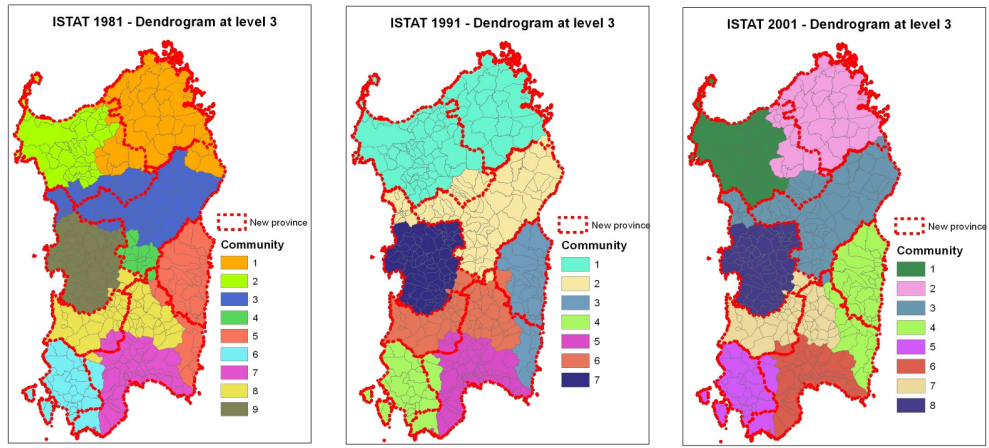

Figure 5 Overlay of layer 3 communities with new provinces in 1981, 1991 and 2001.

The largest inconsistencies are present in the southern area. The communities detected by the Louvain method as populations that heavily commute to Cagliari, do not fit into the boundaries of either the new or the old province of Cagliari. Cagliari, the regional capital, is the most important city in Sardinia. Approximately one-third of the whole population of Sardinia actually lives in the metropolitan area of Cagliari. Community 7 (Figure 5, on the left), community 5 (Figure 5, on the centre) and community 6 (Figure 5, on the right) geographically coincide with the boundaries of the Cagliari metropolitan area. The municipality of Cagliari, the economic center of the island, is an employment dense area. This finding is confirmed by a recent study (Censis, 2008) that identifies the municipality of Cagliari to have one of the highest rates of commuters in Italy. Twenty-eight commuters out of one-thousand citizens live in the metropolitan area of Cagliari. This level of commuting is extremely high especially if compared to the average rate of commuters in some more important municipalities in Italy. On average, in the rest of the nation, nine out of one-thousand citizens commute for work reason (Censis, 2008).

This high level of commuting may be explained looking at the number of buildings erected after the 1991. This information is also confirmed by an increase of 23% of employees in the building construction sector (Censis, 2008).

Both Profili d'Area and the Piano di Rinascita units do not suit well the boundaries provided by the Louvain partitions. Also in the case of Profili d'Area units, the highest similarity emerges for the zone around Oristano (see Figure 6 community 9 on the left, community 7 in the middle and community 8 on the right). This confirms the previous findings. In the province of Oristano there are strong social, economic and historical relationships between the municipalities. Similarly the north area of Sardinia around Sassari shows similarities between the geographical boundaries of Piano di Rinascita and Louvain clusters (Figure 7 community 1 and 2 on the left and right).

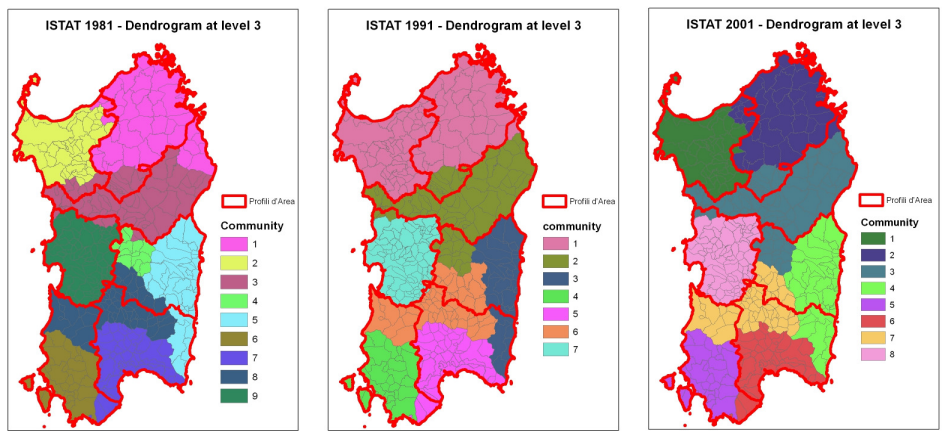

Figure 6 Overlay of layer 3 communities with Profili d'Area in 1981, 1991 and 2001.

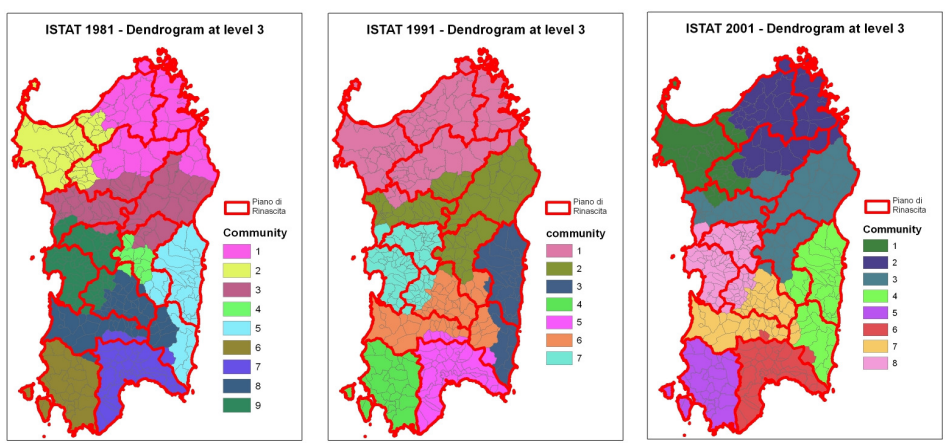

Figure 7 Overlay of layer 3 communities with Piano di Rinascita in 1981, 1991 and 2001

The clustering configuration proposed by the Piano di Rinascita does not overlap the local units shaped by the movements of commuters. With the exception of southern Sardinia, a general high similarity can be detected. The communities 7, 5 and 8 in Figure 7 (in the left, middle and right) shape the boundaries of Piano di Rinascita quite well. Greater similarities can be detected with the communities 6, 4 and 5 in Figure 7 respectively in the left, middle and right side.

We assume that the dissimilarity with Piano di Rinascita is due to the fact that we are comparing information from different periods. The Piano di Rinascita took place during the 1950s while we analyzed data about commuters from 1981 until 2001 using community detection analysis. It is interesting to note that some territorial dynamics and relationships are rooted in the territories from the past.

Figure 8 reports the overlay between level 2 communities of the Louvain method and Comprensori. At first glance, the overlay does not appear to detect a high similarity between the two partitions. A more careful visual analysis shows a hierarchical relation between some units from two different partitions. Some Comprensori include more communities or vice versa. This is exampled in the zone of Oristano, where more Comprensori are included in one Louvain unit. Also in the zone of Tortolì for the year 2001, communities 6, 9, 11 and 12 (Figure 8 on the right) are included in one Comprensorio.

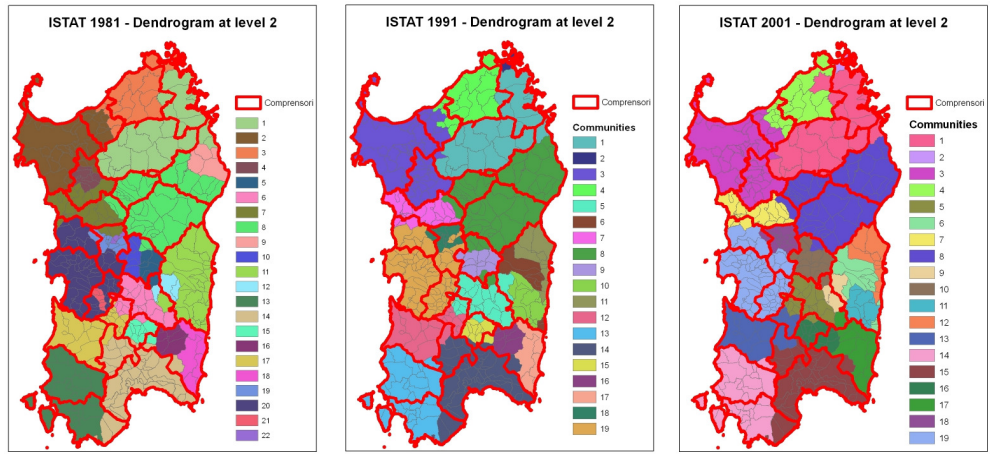

Figure 8 Overlay of layer 2 communities with Comprensori in 1981, 1991 and 2001.

By the application of Louvain method to the commuting network of Sardinia, we have shown that the commuters move among municipalities shaping homogeneous and compact clusters (no disconnected islands were detected). Those relations, studied in a period of 20 years, could suggest a cluster configuration of municipalities as a possible intermediate administrative body comparable to that of the Provinces. Since we applied this methodology to cope with commuter's territorial organization before the administrative process took over, it seems appealing for ex-ante and ex-post planning analysis.

The overlay mapping between the clusters detected by the Louvain method and the other territorial and administrative configurations (province, Comprensori etc) were performed, by comparing the number of clusters in each partition. The number of clusters has sometimes matched the spatial similarity between the partitions (i.e. the province). Thus a quantitative index is needed to substantiate the qualitative results obtained by the overlay mapping.

In the next paragraph we introduce a quantitative method to evaluate the similarity between the partitions already compared in figures from number 4 to 9.

### 5.3 Comparing partitions: a quantitative approach

The Rand index (Rand, 1971) provides a criterion for evaluating partitions from a classification perspective. This index is computed as the ratio of the number of pairs of objects having the same label relationship in two partitions. The Rand index, as it was originally formulated, allows the unique evaluation of hard (crisp or non-fuzzy) clustering partitions. A problem with the Rand index is that the expected value of the Rand index of two random partitions does not take a constant value. Another case for which the Rand index does not provide suitable results is when the data comprise categories that overlap with each other to some degree.

There is a family of other external indices that can be used in order to achieve more accurate results, such as the Adjusted Rand Index as formulated by Hubert and Arabie (1985). The Adjusted Rand index extends the basic Rand index and provides a method able to handle two partitions R and Q of the same data set. The first reference partition R encodes the class labels (i.e. it partitions the dataset into r known classes). Partition Q, in turn, partitions the data into v categories (classes or clusters), and is to be evaluated. According to the above remarks, the adjusted Rand index is defined as follows:

$$\omega_a = a - \frac{\frac{(a+c)(a+b)}{d}}{\frac{(a+c)+(a+b)}{2} - \frac{(a+c)(a+b)}{d}} \tag{3}$$

Where: *a* is the number of pairs of objects belonging to the same class in R and to the same cluster in Q, *b* is the number of pairs of data objects belonging to the same class in R and to different clusters in Q, *c* is the number of pairs of objects belonging to different classes in R and to the same cluster in Q, and *d* is the number of pairs of objects belonging to different classes in R and to different clusters in Q.

The adjusted Rand index gives the degree of agreement between two partitions of a dataset by a value bounded above by 1. A high adjusted Rand index indicates a high level of agreement while a value of 1 suggests a perfect agreement. In the case of random partitions the adjusted Rand index gives a value of 0. In Table 4 the values of adjusted Rand index for the case study are reported. It is worth noting that the values confirm the visual representations of the clusters we have presented in the previous paragraph.

Table 4 Adjusted Rand Index for old, new provinces, Profili d'Area and Piano di Rinascita compared to SMCN layer 3 communities and for Comprensori compared to SMCN layer 2 communities. In the last column we report the average value of Adjusted Rand Index for the 3 years 1981, 1991 and 2001.

|  |  | 1981 | 1991 | 2001 | $<\omega_a>$ |
|---|---|---|---|---|---|
| SMCN Layer 3 communities | New provinces | 0.554 | 0.604 | 0.615 | 0.591 |
|  | Profili d'area | 0.551 | 0.579 | 0.614 | 0.581 |
|  | Old provinces | 0.412 | 0.557 | 0.477 | 0.482 |
|  | Piano rinascita | 0.371 | 0.38 | 0.429 | 0.393 |
| SMCN Layer 2 communities | Comprensori | 0.469 | 0.471 | 0.508 | 0.483 |

The results in Table 4 show that the new provinces and Profili d'Area (values around 0.60) show the best performance in the Adjusted Rand Index analysis. Partitions from the years 1991 and 2001 have the highest similarities. We can postulate that the recent institution of the four new provinces better suits the actual socio-economic dynamics of the Sardinian territory and also suggests that the new fold subdivision was already set into the inner relations between the municipalities even before the official institution of these new provinces. The adjusted Rand index for Comprensori is around 0.50, even if the visual analysis seems to suggest a weak accordance between the partitions. An explanation can be found in the index structure. The index is not able to distinguish the differences between hierarchical units. The hierarchical features found between the two layers reinforce the idea that the Louvain method is able to detect hidden territorial structures.

## 6 Concluding discussion and outlook

In this paper, the authors have discussed the utility to adopt methods able to detect communities in complex networks as tools that could be ordinarily managed by analysts and planners to tack the critical issue of finding optimal homogeneous spatial sub-divisions. The authors have advanced that methods for cluster analysis traditionally adopted by planners and regional analysts may be complemented with tools able to find communities in systems conceived as networks. The arguments are organized around the main hypothesis that intermediate administrative bodies in Sardinia have been designed -even unconsciously- as labour basins constituted by municipalities with similar commuting behaviour. After a state of the art on clustering techniques and community detection methods on networks, the authors proposed a comparison of the communities detected at three Census years in the Sardinian inter-Municipal Commuting Network (SMCN) with relevant administrative subdivisions, which have characterized over the last fifty years the evolution of the intermediate provincial body in Sardinia. The communities have been detected by applying a method studied by Blondel et al (2008) to maximize the modularity function introduced by Newman and Girvan (2004). They also discussed the goodness of fit of various community structures with the relevant sub-divisions in each time period, by adopting the adjusted Rand index, a suitable quantitative measure of the level of similarity of two partitions.

As concluding remarks, first it should be highlighted that the results presented in this paper are overall suggestive, as they show the evolution of the relationship among productive basins, i.e. commuter communities, and the various intermediate bodies. The overall high level of performance of commuter basins especially versus the new provinces, demonstrates that so far local politicians dealing with the definition of the new provinces of Sardinia have taken into account, even unconsciously, the communities' granularity emerging in the SMCN.

Secondly, community detection methods, such as the Louvain method, have proven to be helpful tools for spatial and urban planning. In this case, the method selected and tuned provides results that can be guidance for the analysts, planners and stakeholders in reading, understanding and depicting homogeneous sub-regional systems.

Third, commuting -just one over the many aspects in the design of provincial administrative bodies- still constitutes a determinant factor for shaping provincial administrations boundaries. This may confirm that the phenomenon of commuting contributes to the consolidation of stable relationships among municipalities. Hence, these exchanges –especially when occurring regularly in a long period of time- go far beyond the original work or study movements and imply the strengthening of cultural and landscape cohesive communities.

Besides, the results outlined above are a stimulus to think of future research directions. In this paper, we have demonstrated that in Sardinia commuting relations have historically determined the configuration of provincial administrations. It could be worth testing whether this phenomenon –i.e. the similarity between commuter basins and medium size administrative bodies- occurs in other geographical settings: is it typical of insular setting or is it just geography invariant? A first study may be performed on the island of Sicily -the largest island in the Mediterranean; and other assessments can be developed at the level of regions and states in Europe or in the US.

# 7   References

Anselin L, Getis A, 1992, "Spatial statistical analysis and geographic information systems", *The Annals of Regional Science* **26** 19-33

Anselin L, 1998, "GIS research infrastructure for spatial analysis of real estate markets" *Journal of Housing Research* **9** 113–133

Archibugi F, 1993, "The Urban Mobility Integrated Basin and Its Policy-oriented Identification: A Prerequisite of Rationality for Planning of Urban Transport" in *Proceedings of the 32nd Scientific Meeting of SIEDS*, Taormina, Italy, May 1993

Archibugi F, 1994, "Urbanistica ed ecologia: quale rapporto? Alcune considerazioni sulla definizione di un metodo integrativo" in *Proceedings of the 8th Congress of the Association of European Schools of Planning*, Istambul, Turkey, 24-27 August 1994

Albert R, Barabási AL, 2002, "Statistical mechanics of complex networks", *Rev. Mod. Phys.* **74**, 47-97

Barabàsi AL, Albert R, 1999, "Emergence of scaling in random networks", Science **286** 509-512

Barrat A, Barthélemy M, Pastor-Satorras R, Vespignani A, 2004, "The architecture of complex weighted networks" *Proceedings of The National Academy of Sciences* **11** 3747-3752

Barthélemy M, Barrat A, Pastor-Satorras R, Vespignani V, 2005, "Characterization and modelling of weighted networks" *Physica A* **346**, 34-43

Barthélemy M, Flammini A, 2008, "Modeling Urban Street Patterns" *Physical Review Letters* **100** (13) 138702

Barthélemy M, Flammini A, 2009, "Co-evolution of density and topology in a simple model of city formation" *Networks and Spatial Economics* **9** (3) 401-425

Batty M, Sammons R, 1978, "On searching for the most informative spatial pattern" *Environment and Planning A* **10**, 747-779

Batty M, Xie Y, 1994, "Modeling Inside GIS: Part I: Model Structures, Exploratory Spatial Data Analysis, and Aggregation" *International Journal of Geographical Information Systems* **8**(3) 291-307

Batty M., 2001, "Cities as small worlds, Editorial" *Environment and Planning B: Planning and Design* **28** 637-638

Batty M., 2008a "Cities as Complex Systems: Scaling, Interactions, Networks, Dynamics and Morphologies" *UCL Working Paper Series*

Batty M., 2008b, "The size, scale, and shape of cities" Science 319(5864) 769-771

Berry BJL, 1964, "Approaches to regional analysis: a synthesis" *Annals of the Association of the American Geographers* **54**, 2-11

Berry BJL, 1966, "Reflections on the functional economic areas", in *Research and Education for regional and area development* Eds WR Maki, BJL Berry, (Iowa State University Press, Ames USA), pp 56-64

Berry BJL, 1967, "Grouping and regionalization: An approach to the problem using multivariate analysis" in Garrison W.L. and D.F. Marble (eds.) Quantitative Geography. Part I: Economic and cultural topics. *Northwestern University Studies in Geography* **13** 219-251

Berry BJL, 1968, "A synthesis of formal and functional regions using a general field theory of behavior", in *Spatial analysis: a reader in statistical geography* Eds BJL Berry and DF Marble (Prentice-Hall, Englewood Cliffs, New Jersey USA), pp 419-428

Blondel VD, Guillaume J, Lambiotte R, Lefebvre E, 2008, "Fast unfolding of communites in large networks" *Journal of Statistical Mechanics: Theory and Experiment* **10** 10008-10020

Campagna M, Caschili S, Chessa A, De Montis A, Deplano G, 2007, "Modeling commuters dynamics as a complex network: the influence of space", in *Proceedings of the 10th International Conference on Computers in Urban Planning and Urban Management (CUPUM)*, Iguassu Falls, Brasil, July, 11-13 2007

Censis, 2008 *Pendolari d'Italia: scenari e strategie* (Franco Angeli, Milano)

Crucitti P, Latora V, Porta S, 2006, "Centrality measures in spatial networks of urban streets", *Physical Review E* **73** 036125

Davis WM, 1924, "The Progress of Geography in the United States" *Annals of the Association of American Geographers* **14** 159-215

Davies WKD, 1980, "Higher order factor analysis and functional regionalization: A case study in south Wales 1971". *Environment and Planning A* **12** 685-701

De Montis A, 1997, "Le intenzioni della pianificazione del territorio nell'esame degli strumenti di programmazione della "area vasta": il caso della Sardegna", in *Proceeding of 18th Italian Conference of Regional Science* volume 1, 8-11 October 1997, Siracusa, Italy, pp 323-333, in Italian

De Montis A, Barthélemy M, Chessa A, Vespignani A, 2007, "The structure of interurban traffic: a weighted network analysis" *Environment and Planning B: Planning and Design* **34**(5) 905-924

De Montis A, Campagna M, Caschili S, Chessa A, Deplano G, 2010, "Modelling commuting systems through a complex network analysis: a Study of the Italian islands of Sardinia and Sicily" *Journal of Transport and Land Use* **2**(3-4) 39-55

De Montis A, Caschili S, Chessa A, 2011, "Time Evolution of Complex Networks: Commuting Systems in Insular Italy" *Journal of Geographical Systems* **13**(1) 49-65

Erdös P, Rényi A, 1959, "On Random Graphs I," *Publicationes Mathematicae* **6** 290-297

Erdös P, Rényi P, 1960, "On the evolution of random graphs" *Publ. Math. Inst. Hung. Acad.* **5** 17-60

Fischer MM, 1979, "Regional Taxonomy: A Comparison of some Hierarchic and Non-Hierarchic Strategies" *Regional Science and Urban Economics* **10**, 503-537

Fischer MM, 1982, "Eine Methodologie der Regionaltaxonomie: Probleme und Verfahren der Klassifikation und Regionalisierung in der Geographie und Regionalforschung" *Bremer Beiträge zur Geographie und Raumplanung*, Presse und Informationsamt, Universität Bremen, Bremen

Fischer MM, 1984, "Regional taxonomy: some reflections on the state of the art" in *Territorial planning and information system* Ed F Clemente (Franco Angeli, Milano) pp 753-776

Flake GW, Lawrence S, Lee Giles C, Coetzee FM, 2002, "Self-Organization and Identification of Web Communities" *IEEE Computer* **35**(3) 66–71

Fortunato S, Castellano C, 2009, "Community structure in graphs", *Springer´s Encyclopedia of Complexity and System Science*

Fortunato S, 2009, "Community detection in graphs" *Physics Reports* **486** 75-174

Fox KA, 1967, "Functional Economic Areas and consolidated urban regions of the United States" *Social Science Research Council* ITEMS 21

Fox KA, 1974, "Elements of an Operational System. II: Cities and Regions" *Social Indicators and Social Theory, Elements for an Operational System*, Ed. in KA Fox (Wiley, New York) Chapter XII

Garrison WL, Marble DF, 1963, "Factor analytic study of the connectivity of a transportation network" *Papers and Proceedings, Regional Science Association* **12**, 231-238

Gastner MT, Newman MEJ, 2006, "The spatial structure of networks" *The European Physical Journal* B **49** 247-252.

Girvan M, Newman M E J, 2002, "Community structure in social and biological networks" *Proceedings of The National Academy of Sciences* **99** 7821–7826

Goodchild MF, Haining RP, Wise S, 1992, "Integrating GIS and spatial analysis- Problems and possibilities" *International Journal of Geographical Information Systems* **6** 407-423

Gorman SP, Kulkarni R, 2004, "Spatial small worlds: new geographic patterns for an information economy" *Environment and Planning B: Planning and Design* **31** 273-296

Guimerà R, Amaral LAN, 2005, "Functional cartography of complex metabolic networks" *Nature* **433** 895-900

Hubert LJ, Arabie P, 1985, "Comparing partitions" *Journal of Classification* **2** 193-218

Jiang B, Claramunt C, 2004, "Topological analysis of urban street networks" *Environment and Planning B: Planning and Design* **31** 151-162

Krause AE, Frank KA, Mason DM, Ulanowicz RE, Taylor WW, 2003, "Compartments exposed in food-web structure" *Nature* **426** 282–285

Iacovacci G, 1995, "Sull'utilizzo del metodo delle c-medie sfocato per la classificazione dei comuni italiani a seconda del grado di urbanità e ruralità" Statistica Applicata **7**(1) 33-48

Latora V, Marchiori M, 2003, "Economic small-world behavior in weighted networks" *The European Physical Journal B* **32** 249-263

Leung, Y, 1988 *Spatial Analysis and Planning under Imprecision* (North-Holland, Amsterdam)

Lusseau D, Newman MEJ, 2004, "Identifying the role that animals play in their social networks" *Proceedings of the Royal Society of London B* **271** S477–S481

Masser I, Brown PJB, 1975, "Hierarchical aggregation procedures for interaction data" *Environment and Planning A* **7** 509-523

Milgram S, 1967, "*The small-world problem*", Psychology Today **1** 61-67

Newman MEJ, 2003, "Structure and function of complex networks*" SIAM review* **45** 167-256

Newman MEJ, Girvan M, 2004, "Finding and evaluating community structure in networks" *Phys. Rev. E* **69** 026113-026128

Noronha VT, Goodchild MF, 1992, "Modeling Interregional Interaction: Implications for Defining Functional Regions" *Annals of the Association of American Geographers* **82**, 86-102

Palla G, Derenyi I, Farkas I, Vicsek T, 2005, "Uncovering the overlapping community structure of complex networks in nature and society" *Nature* **435** 814-818

Pastor-Satorras R, Vespignani A, 2004 *Evolution and Structure of the Internet* (Cambridge University Press, Cambridge, USA)

Patuelli R, Reggiani A, Gorman SP, Nijkamp P, Bade FJ, 2007, "Network Analysis of Commuting Flows: A Comparative Static Approach to German Data" *Networks and Spatial Economics* **7**(4) 315-331

Pimm SL, 1979, "The structure of food webs" *Theoretical Population Biology* **16** 144–158

Rand W.M, 1971, "Objective criteria for the evaluation of clustering methods" *Journal of the American Statistical Association* **66**: 846–850

Regione Autonoma della Sardegna (RAS), 2001, Legge Regionale 12 luglio 2001, n. 9, Istituzione delle province di Carbonia-Iglesias, del Medio Campidano, dell'Ogliastra e di Olbia-Tempio (Regional law on "Institution of the provinces of Carbonia-Iglesias, Medio Campidano, Ogliastra, and Olbia-Tempio"), website: http://consiglio.regione.sardegna.it/sito/Manuale%20consiliare/ManTomoI.asp, accessed on March, 26 2007

Regione Autonoma della Sardegna (RAS), 2003, Legge Regionale 13 ottobre 2003, n. 10, Ridelimitazione delle circoscrizioni provinciali (Regional law on "Ri-definition of the provincial electoral districts"), website: http://consiglio.regione.sardegna.it/sito/Manuale%20consiliare/ManTomoI.asp, accessed on March, 26 2007

Rolland-May, 1985, "Fuzzy geografical space: algorithms of fuzzy and application to fuzzy regionalization" *Sistemi Urbani* **3** 237-257

Rushton, G, 2003, "Public health, GIS, and spatial analytic tools" *Annual Review of Public Health* **24**(3) 43-56

Schintler LA, Gorman SP, Reggiani A, Patuelli R, Gillespie A, Nijkamp P, Rutherford J, 2005, "Complex Network Phenomena in Telecommunication Systems" *Networks and Spatial Economics* **4** 351-370

Semple RK, Youngmann CE, Zeller RE, 1972, "Economic regionalization and information theory: An Ohio example" *Discussion Paper* **28**, Department of Geography, Ohio State University, Columbus

Song C, Havlin S, Makse HA, 2005, "Self-Similarity of Complex Networks" *Nature* **433** 392-395

Spence NA, Taylor PJ, 1970, "Quantitative methods of regional taxonomy" *Progress in Geography* **2**, l-64

Symanzik J, Majure J, Cook D, 1996, "Dynamic graphics in a GIS: a bidirectional link between ArcView 2.0 and Xgobi" *Computing Science & Statistics* **27** 299-303

Watts DJ, Strogatz SH, 1998, "Collective dynamics of 'small-world' networks" *Nature*, **393**, 440-442

Wheeler JO, 1986, "Notes on the rise of the area studies tradition in U.S. geography, 1910–1929" *The Professional Geographer* **38**, 53-61

Zadeh LA, 1977, "Fuzzy set and their application to pattern classification and clustering", in Classification and Clustering Ed J Van Ryzin (Accademic Press, New York) pp 251-299